\input harvmac
 
\input epsf
\overfullrule=0pt  
\parindent=0pt

\ifx\epsfbox\UnDeFiNeD\message{(NO epsf.tex, FIGURES WILL BE IGNORED)}
\def\figin#1{\vskip2in}
\else\message{(FIGURES WILL BE INCLUDED)}\def\figin#1{#1}\fi

\def\ifig#1#2#3{\xdef#1{fig.~\the\figno}
\goodbreak\figin{\centerline{#3}}%
\smallskip\centerline{\vbox{\baselineskip12pt
\advance\hsize by -1truein\noindent{\bf
Fig.~\the\figno:} #2}}
\bigskip\global\advance\figno by1}

\newwrite\ffile\global\newcount\figno \global\figno=1
\def\fig{fig.~\the\figno\nfig}
\def\nfig#1{\xdef#1{fig.~\the\figno}%
\writedef{#1\leftbracket fig.\noexpand~\the\figno}%
\ifnum\figno=1\immediate\openout\ffile=figs.tmp\fi\chardef\wfile=
\ffile%
\immediate\write\ffile{\noexpand\medskip\noexpand\item{Fig.\
\the\figno. }
\reflabeL{#1\hskip.55in}\pctsign}\global\advance\figno by1\findarg}

\def\xxx#1 {{hep-th/#1}}
\def\lr { \lref}
\def\npb#1(#2)#3 { Nucl. Phys. {\bf B#1} (#2) #3 }
\def\rep#1(#2)#3 { Phys. Rept.{\bf #1} (#2) #3 }
\def\plb#1(#2)#3{Phys. Lett. {\bf #1B} (#2) #3}
\def\prl#1(#2)#3{Phys. Rev. Lett.{\bf #1} (#2) #3}
\def\physrev#1(#2)#3{Phys. Rev. {\bf D#1} (#2) #3}
\def\ap#1(#2)#3{Ann. Phys. {\bf #1} (#2) #3}
\def\rmp#1(#2)#3{Rev. Mod. Phys. {\bf #1} (#2) #3}
\def\cmp#1(#2)#3{Comm. Math. Phys. {\bf #1} (#2) #3}
\def\mpl#1(#2)#3{Mod. Phys. Lett. {\bf #1} (#2) #3}
\def\ijmp#1(#2)#3{Int. J. Mod. Phys. {\bf A#1} (#2) #3}
\def\mpla#1(#2)#3{Mod. Phys. Lett. {\bf A#1} (#2) #3}
\def\jhep#1(#2)#3{JHEP {\bf  #1} (#2) #3}

\parindent 25pt
\overfullrule=0pt
\tolerance=10000

\def\half{{\textstyle {1 \over 2}}}
\def\quart{{\textstyle {1 \over 4}}}

\sequentialequations


\lr\costas{C. Bachas, C. Fabre, E. Kiritsis, N.A. Obers and P.
Vanhove, {\it Heterotic / type I duality and D-brane instantons},
\npb509(1998)33, hep-th/9707126.} 
\lr\kiritsis{E. Kiritsis and
N.A. Obers, {\it  Heterotic type I duality in D $<$ 10-dimensions,
threshold corrections and D instantons}, \jhep10(1997)004,
hep-th/9709058.}
 \lr\polchinskia{J. Polchinski and E. Witten, {\it
Evidence for heterotic - type I string duality}, \npb460(1996)525,
\xxx9510169.}

 \lr\bgs{C. Bachas, M.B.
Green and A. Schwimmer, {\it (8,0) quantum mechanics and symmetry
enhancement in type I' superstrings}, \jhep01(1998)006,
hep-th/9712086.}
 \lr\kachru{Kachru and Silverstein, {\it On gauge
bosons in the matrix model approach to M theory}, \plb396(1997)70,
\xxx9612162.}
 
\lr\dixon{L. Dixon,
V. Kaplunovski and J. Louis, {\it Moduli dependence of string loop
corrections to gauge coupling constants}, \npb355(1991)649.}

\lr\lerche{W. Lerche and S. Stieberger,
{\it Prepotential, mirror map and F theory on K3}, \xxx9711107; W.
Lerche, S. Stieberger
  and N. Warner, {\it Quartic gauge couplings from K3 geometry},  \xxx9811228.}

 \lr\ellis{J. Ellis, P. Jetzer, L. Mizrachi, {\it One loop string corrections
 to the effective field theory}, \npb303(1988)1.}

 \lr\ellgenus{W. Lerche, B. Nilsson and A. Schellekens, {\it Heterotic String
  Loop Calculation Of The Anomaly Cancelling Term}, \npb289(1987)609;
  W. Lerche, B. Nilsson, A. Schellekens and N. Warner, {\it Anomaly Cancelling
  Terms From The Elliptic Genus},
  \npb299(1988)91.}
  \lr\lercheb{W. Lerche, {\it Elliptic Index And Superstring Effective Actions}, \npb308(1988)102.}


\lr\senor{A.~Sen,
{\it F theory and orientifolds}, 
Nucl. Phys. {\bf B475} (1996) 562,  
hep-th/9605150.}
\lr\probe{T.~Banks, M.R.~Douglas and N.~Seiberg,
{\it Probing F theory with branes}, 
Phys. Lett. {\bf B387} (1996) 278 
hep-th/9605199; M.R.~Douglas, D.A.~Lowe and J.H.~Schwarz,
{\it Probing F theory with multiple branes}, 
Phys. Lett. {\bf B394} (1997) 297,
hep-th/9612062.}
\lr\aharony{O.~Aharony, A.~Fayyazuddin and J.~Maldacena,
{\it The Large N limit of N=2, N=1 field theories from three-branes in
  F theory}, JHEP {\bf 07} (1998) 013, 
hep-th/9806159.}
\lr\spal{A.~Fayyazuddin and M.~Spalinski,
{\it Large N superconformal gauge theories and supergravity
  orientifolds}, 
Nucl. Phys. {\bf B535} (1998)  219,
hep-th/9805096.}
\lr\stephanb{P. Mayr and S. Stieberger, {\it Threshold corrections to
gauge couplings in orbifold compactifications}, \npb407(1993)725,
\xxx9303017.}
\lr\juan{J. Maldacena, {\it The Large N limit of superconformal field
  theories and supergravity}, 
Adv. Theor. Math. Phys. {\bf 2} (1998)  231,  
hep-th/9711200.}
\lr\adscft{S.S.~Gubser, I.R.~Klebanov and A.M.~Polyakov,
{\it Gauge theory correlators from noncritical string theory}, 
Phys. Lett. {\bf B428} (1998)  105 
hep-th/9802109; E.~Witten,
{\it  Anti-de Sitter space and holography}, 
Adv. Theor. Math. Phys. {\bf 2} (1998) 253,  
hep-th/9802150.}
\lr\onelothr{K.~Foerger and S.~Stieberger,
{\it Higher derivative couplings and heterotic type I duality in eight-
 dimensions}, hep-th/9901020; K.~Foerger,
{\it On heterotic / type I duality in d = 8}, 
hep-th/9812154.}
\lr\onelothrb{M. Bianchi, E. Gava, F. Morales and K.S. Narain, {\it D
strings in unconventional type I vacuum configurations}, \xxx9811013.}
\lr\gutb{M.~Gutperle,
{\it A Note on heterotic / type I-prime duality and D0 brane quantum
  mechanics}, JHEP {\bf 05} (1999) 007, hep-th/9903010.}
\lr\guta{M.B. Green and M. Gutperle,
{\it Effects of D instantons},
\npb498(1997)195, hep-th/9701093.}
\lr\freedman{D.Z.~Freedman, S.D.~Mathur, A.~Matusis and L.~Rastelli,
{\it Correlation functions in the CFT(d) / AdS(d+1) correspondence},
Nucl. Phys. {\bf B546} (1999)  96, 
hep-th/9804058.}
\lr\pierre{I.K.~Kostov and P.~Vanhove,
{\it Matrix string partition functions},
Phys. Lett. {\bf B444} (1998) 196, 
hep-th/9809130; P.~Vanhove,
{\it D instantons and matrix models},
hep-th/9903050.}
\lr\osborn{H.~Osborn,
{\it Semiclassical Functional Integrals For Selfdual Gauge Fields}, 
Ann. Phys. {\bf 135} (1981) 373.}
\lr\orbf{L.~Dixon, D.~Friedan, E.~Martinec and S.~Shenker,
{\it  Conformal Field Theory Of Orbifolds}, 
Nucl. Phys. {\bf B282} (1987) 13.}
\lr\banksgreen{T.~Banks and M.B.~Green,
{\it Nonperturbative effects in AdS in five-dimensions x S**5 string
theory and d = 4 SUSY Yang-Mills}, 
JHEP {\bf 05} (1998) 002, 
hep-th/9804170.}
\lr\bianchi{M.~Bianchi, M.B.~Green, S.~Kovacs and G.~Rossi,
{\it Instantons in supersymmetric Yang-Mills and D instantons in IIB
superstring theory}, 
JHEP {\bf 08} (1998) 013, 
hep-th/9807033.}
\lr\mattis{N.~Dorey, T.J.~Hollowood, V.V.~Khoze, M.P.~Mattis and S.~Vandoren,
{\it Multiinstantons and Maldacena's conjecture}, 
hep-th/9810243; {\it Multi-instanton calculus and the AdS / CFT
correspondence in N=4 superconformal field theory},
hep-th/9901128.}
\lr\polchinski{J.~Polchinski,
{\it Tasi lectures on D-branes},
hep-th/9611050.}
\lr\zeroeight{S.~Kachru and E.~Silverstein, {\it On gauge
bosons in the matrix model approach to M theory}, \plb396(1997)70,
\xxx9612162;  D.~Lowe, {\it Bound states of type I-prime D particles
and enhanced gauge symmetry}, \npb501(1997)134, \xxx9702006;  U.H.~Danielsson and G.~Ferretti,
{ \it The Heterotic life of the D particle}, \ijmp12(1997)4581,
\xxx9610082. }
\lr\pgg{G.W.~Gibbons, M.B.~Green and M.J.~Perry,
{\it Instantons and seven-branes in type IIB superstring theory},
Phys. Lett. {\bf B370} (1996)  37,
hep-th/9511080.}
\lr\seibwit{N.~Seiberg and E.~Witten, {\it Monopoles, duality and chiral symmetry breaking in N=2
  supersymmetric QCD},  Nucl. Phys. {\bf B431} (1994)  484 
hep-th/9408099.}
\lr\mattisb{N.~Dorey, V.V.~Khoze and M.P.~Mattis,
{\it On N=2 supersymmetric QCD with four flavors},
Nucl. Phys. {\bf B492} (1997)  607, 
hep-th/9611016.}
\lr\friedan{D.~Friedan, E.~Martinec and S.~Shenker,
{\it Conformal Invariance, Supersymmetry And String Theory},
Nucl. Phys. {\bf B271} (1986) 93.}
\lr\gutc{M.B.~Green and M.~Gutperle,
{\it D Particle bound states and the D instanton measure},
JHEP {\bf 01} (1998)  005, 
hep-th/9711107; {\it D instanton partition functions},
Phys. Rev. {\bf D58} (1998) 046007 
hep-th/9804123.}
\lr\nikita{G.~Moore, N.~Nekrasov and S.~Shatashvili,
{\it D particle bound states and generalized instantons},
hep-th/9803265.}
\lr\bbg{C.P.~Bachas, P.~Bain and M.B.~Green,
{\it Curvature terms in D-brane actions and their M theory origin},
JHEP {\bf 05} (1999)  011, 
hep-th/9903210.}
\lr\gava{E.~Gava, J.F.~Morales, K.S.~Narain and G.~Thompson,
{\it Bound states of type I D strings}, 
Nucl. Phys. {\bf B528} (1998)  95, 
hep-th/9801128.}
\lr\gavab{E.~Gava, A.~Hammou, J.F.~Morales and K.S.~Narain,
{\it On the perturbative corrections around D string instantons},
JHEP {\bf 03}  (1999) 023, 
hep-th/9902202.}
 \noblackbox 
\baselineskip 18pt plus 2pt minus 2pt
\Title{\vbox{\baselineskip12pt
\hbox{hep-th/9905173}
\hbox{PUPT-1866}
}}
{\vbox{
\centerline{ Heterotic/type I duality, D-instantons  }
 \centerline{and a $N=2$ ADS/CFT correspondence} }}
\centerline{Michael Gutperle\foot{email:
gutperle@feynman.princeton.edu}} 
\medskip
\centerline{Department of Physics, Princeton University, Princeton NJ
08554, USA} \bigskip 

\medskip
\centerline{{\bf Abstract }}
\noblackbox
D-instanton effects are studied for the  IIB   orientifold $T^2/(-1)^{F_L}
\Omega I$ of Sen  using type I/heterotic duality. An exact one
loop threshold calculation  of $t_8 \tr F^4$ and $t_8(\tr F^2)^2$ terms for the
heterotic string on $T^2$ with Wilson lines breaking $SO(32)$ to $SO(8)^4$
is related  to D-instanton induced terms in the worldvolume of D7 branes in the
orientifold. Introducing D3 branes and using the AdS/CFT
correspondence in this case, these terms are used to calculate
Yang-Mills instanton contributions to four point functions of the
large $N_c$ limit of $N=2$ $USp(2N_c)$ SYM with four fundamental and
one antisymmetric tensor hypermultiplets. 
\baselineskip 16pt plus 2pt minus 2pt

\Date{May 1999}
\vfill\eject

\newsec{Introduction}
One of the simplest examples of F-theory on $K3$ was introduced
by Sen \senor. This  compactification is
realized as a perturbative IIB orientifold $T^2/(-1)^{F_L}\Omega
I$, where $\Omega$ is the worldsheet orientation reversal, and $I$ is
the inversion $z\to -z$ on the torus.  Orientifold
seven planes are  
located at the four  fixed point of the inversion $I$ on $T^2$. In addition four
D7 branes are on top of the each of the orientifold seven planes. The 
sources for  the dilaton and the Ramond-Ramond scalar  coming from the D7
branes and the 
 orientifold seven planes cancel locally and hence the dilaton is constant
 over the base  $T^2/I$.  The D7 branes  lead to  an
 enhanced gauge symmetry $SO(8)^4$. Each $SO(8)$ factor
can be associated with  four D7 branes on top of one of the four
orientifold planes.

When  a probe of $N_c$  parallel  D3 branes
\probe\ is moving in the vicinity of 
 one of the orientifold planes the low energy field theory on the
probe is given by a $N=2$, $USp(2N_c)$ gauge
theory with four hypermultiplets transforming in the fundamental
representation   and one hypermultiplet transforming in
the second rank antisymmetric tensor representation of $USp(2N_c)$
 respectively. 

Such a theory has an exactly vanishing beta function and defines a
conformal field theory for any $N_c$  when  the expectation values for
the scalars 
in the hypermultiplets 
are zero. In \spal\aharony\ the large
$N_c$ limit of the probe field theory  was considered using
the AdS/CFT correspondence of Maldacena \juan. The conformal
field theory is related to IIB superstring theory compactified on an
orientifold $AdS_5\times S_5/Z_2$. 

Sen's IIB orientifold is related by T-duality to   type I theory
compactified on $T^2$ with Wilson lines turned on. This theory is in
turn mapped by S-duality  to the heterotic string theory on $T^2$ with
Wilson lines 
breaking $SO(32)$ to $SO(8)^4$. In the following this
chain of dualities is used  to relate a one loop calculation of four derivative
threshold correction on the heterotic side to D-instanton induced four
derivative terms localized in the worldvolume of the seven branes.

In the Maldacena limit the seven branes wrap a $S_3$ in  $S_5/Z_2$ and fill
$AdS_5$. Therefore terms in the worldvolume action of the seven brane induce
vertices in the $AdS_5$  which contribute to correlation function in
the CFT using the prescription introduced in \adscft.

\newsec{Two torus compactification with Wilson lines}

We are interested in the $SO(32)$ heterotic string compactified on a
two torus. The K\"ahler
and complex structure modulus of $T^2$ are denoted $T$ and $U$ respectively. The $SO(32)$
gauge symmetry will be broken  to $SO(8)^4$ by introducing Wilson
lines on the two torus of the following form
\eqn\wilsonl{Y_i^1=(0^4,0^4,\half^4,\half^4),\quad Y_i^2=(0^4,\half^4,0^4,\half^4). } This
choice of Wilson lines on the heterotic side is dual to the IIB orientifold 
$T^2/(-1)^{F_L}\Omega I$.  Each $SO(8)$ factor
can be associated with  four D7 branes on top of one of the four
orientifold planes located at the fixed points of $T^2/I$.

The one loop heterotic thresholds discussed \ellis\lercheb\  are related by supersymmetry to anomaly canceling
terms \ellgenus\ and are   presumably exact at one loop. The one loop
integrands  involved are almost holomorphic since only BPS-states
propagate in the loop and  are related to the 'elliptic genus' \ellgenus.

 For simplicity  
a square torus with radii $R_1,R_2$ will be considered here. The
K\"ahler and complex 
structure moduli are then given  by 
\eqn\kahlcom{T=
B^{NS}_{12}+i R_1R_2,\quad U= i {R_2\over R_1}.}
 Under heterotic-type I duality the coupling constants, metric and AST
 field  are 
related by \eqn\hettypi{\lambda_{het}=1/\lambda_I,\quad  \lambda^I
  g_{\mu\nu}^{het}=g^I_{\mu\nu},\quad B_{\mu\nu}^{het}=B_{\mu\nu}^I.}
Two T-dualities invert the radii $R_i\to1/R_i$,
 $i=1,2$. Under this operation   type I gets
mapped to the  type IIB orientifold of Sen \senor, where the radii and
other fields are related  by
\eqn\tdual{R_1^I =1/ R_1, \quad R_2^I =1/ R_2,
\quad  R_1^IR_2^I
  \lambda=\lambda^I,\quad B^{RR,I}_{12}=\chi. }
 Here the superscript $I$ denotes type I fields and no superscripts
 denotes fields of the IIB orientifold. $\chi$ denotes the RR
 scalar. Hence the heterotic moduli T and U get 
mapped to the following fields in the IIB orientifold 
\eqn\typeonep{T\to \tau = \chi +i {1\over \lambda},\quad U\to U=
i{R_1\over
  R_2}.}
It follows from \typeonep\ that the heterotic modulus $T$ can be
interpreted as the D-instanton action on the orientifold side.
   Hence worldsheet
instantons on the heterotic side (i.e. fundamental string worldsheets
wrapping $T^2$) which are weighted by a factor $\exp(-2\pi  T_2)$  will be
identified on the orientifold side with D-instantons which are weighted by a factor $\exp(-2\pi  /\lambda)$ .

In the following we will be interested in 
the  one
loop thresholds of the form  $t_8\tr(F^4)_{1}$ and
$t_8(\tr(F^2)_{1})^2$ in the presence of the   Wilson line \wilsonl. The
subscript $_1$ on the field  strength 
indicates that the trace is taken over the first $SO(8)$ factor
and without loss of generality can be   associated with the four D7
branes at one of the 
orientifold planes. 

These  calculations were first performed by Lerche
and Stieberger in \lerche\ 
using methods developed in \stephanb. In particular the thresholds in
eq (6) can be read off from eq (2.22) in  \lerche. Other one loop
calculations 
of thresholds 
in related contexts can be found in \onelothr\onelothrb\gavab.

 For the interested reader the 
details of the threshold calculation   are presented in the appendix  using 
a somewhat
different method (also employed in \gutb\  for $SO(16)^2$). Only the
result given in  eq. B.8 and B.16 will be important for the main
arguments of the paper.
 
 It turns out that the $U$ and $T$
dependence of the result is decoupled and the $U$ dependent terms come
from the trivial and degenerate orbits, whereas the $T$ dependent
terms are given by the non degenerate orbits.\foot{The parameter
  $U$ is related to the conformal cross-ratio of the position of the
  seven branes in $T^2/I$ } Since we are interested in
D-instanton induced terms on the worldvolume of the seven branes only
the contributions of the non degenerate orbits are considered here.
 The results from appendix B are  (see eq. B.8 and B.16):

\eqn\seveninst{\eqalign{I_7&=\int d^8x\; t_8
\tr(F^4)\sum_N\Big({1\over 2}\sum_{N|m}{1\over m} e^{2\pi i
2NT}-{1\over 2}\sum_{N|m}{1\over m} e^{2\pi i 4NT}\Big)\cr &+\int d^8x
\;t_8 (\tr(F^2))^2 \sum_N\Big({1\over 4}\sum_{N|m}{1\over m} e^{2\pi i
2NT}-{1\over 8}\sum_{N|m}{1\over m} e^{2\pi i 4NT}\Big)+cc.}}
Using \typeonep\ together with the fact that since it is a one loop
amplitude,  \seveninst\ is
independent of the heterotic string coupling (which in turn is related
to the volume of $T^2/I$). It is then easy to see that \seveninst\
corresponds to D-instanton induced terms localized on the worldvolume
of the seven branes. In \seveninst\ only the parity even contribution
was calculated, the parity odd contribution which contains an eight
dimensional epsilon tensor $\epsilon_8$ instead of $t_8$ can be
calculated in the same way and the only difference is  a minus sign instead of a plus
sign  between the holomorphic
(instanton) and and antiholomorphic (anti-instanton) part in \seveninst.

Note that  only even
 number of instantons contribute to the
 four derivative terms in \seveninst. 
The structure of these terms is simpler than the one found in related
instanton induced threshold terms like $t_8t_8  R^4$ terms in IIB
 \guta. 
  The $t_8\tr(F)^4$ and the  $t_8(\tr(F)^2)^2$ thresholds  in \seveninst\
 do not receive any corrections as a power series  $T_2$, which on
 the IIB orientifold side are
 interpreted as perturbative corrections to the D-instanton (as in
 \guta). One possible
 interpretation is that there is an exact cancellation of bosonic and
 fermionic fluctuations in this case \costas\pierre\ in analogy to
 cancellations of 
 bosonic and fermionic determinants in the background of
 supersymmetric Yang-Mills
 instantons \osborn.

\newsec{Instanton induced interactions}
A D-instanton is a D $p=-1$ brane, which means that the worldvolume is a
point in space time and the bosonic coordinates of open strings ending
on the D-instanton $X^\mu$ satisfy Dirichlet boundary conditions for
all directions 
$\mu=0,\cdots,9$. The presence of an orientifold plane
changes the representation of  Chan-Paton factors labeling open
strings stretched between  several D-instantons. The  worldvolume of
the orientifold
seven plane will be chosen to  lie  in the $\mu =0,\cdots 7$ directions. The
Chan-Paton factors for $k$  D-instanton fall in representations of
$SO(k)$.
The vertex operator for the 'massless' bosonic states is given by
\eqn\vertrep{\eqalign{V^i &= M_{(IJ)} \partial_n X^i,\quad  i=0,\cdots,7\cr
V^a &= M_{[IJ]} \partial_n X^a,\quad  a=8,9.}}
The fact that the longitudinal vertex $V^i$ and transverse vertex
$V^a$ transform as second rank symmetric and antisymmetric tensors of
$SO(k)$ respectively is determined by the consistency of action of the orientifold
$(-1)^F_l\Omega I$ on the open string vertices and the Chan-Paton
factors \polchinski.

In addition there are fermionic collective coordinates which  like
their bosonic partners in \vertrep\ come in two representations.
\eqn\fermcoll{\eqalign{ V^a &= M_{(IJ)} e^{-\half \phi} S^{\alpha}
e^{-i\half H_5},\cr
 V^{\dot a} &= M_{[IJ]} e^{-\half \phi} S^{\dot \alpha}
e^{+i\half H_5}. }}
Here $\phi$ is the bosonized superghost and the fermionic vertices are
in the  $-1/2$ picture.  A $SO(10)$ spin field $\Sigma^a$ of definite
chirality is
decomposed as $S^{\alpha}
e^{-i\half H_5}$ and $S^{\dot \alpha}
e^{+i\half H_5}$.  Here  $S^{\alpha}, S^{\dot \alpha}$ are $SO(8)$
spin fields  of opposite chirality and a
$SO(2)$ spin field is bosonized $e^{\pm
i\half H_5}$  
using the free 
boson $H_5$ in
\fermcoll.  We identify the $SO(8)$ directions with the worldvolume
directions of the orientifold seven plane. The sign of $e^{\pm
i\half H_5}$ determines the parity under the orientifold projection
and hence the representation of the $SO(k)$ Chan-Paton factors in \fermcoll.

The presence of D7 branes introduces further collective coordinates,
corresponding to stretched strings between the D7 brane and the
D-instanton. The analysis of such open string states is quite subtle,
since there are eight ND directions, like in the D0-D8 system \zeroeight\bgs. An
analysis of the normal ordering constants for the Virasoro constraint
of this system shows that all states in the NS sector are massive and
that there is a single physical state in the Ramond sector (after GSO
projection) which can be interpreted as a fermionic stretched string.

The vertex operator for this  state when inserted on the boundary of a
worldsheet changes the boundary condition
from Neumann (7-brane) to Dirichlet (D-instanton) for $X^i, i=0,\cdots,7$.
Such boundary condition changing operators can be  constructed using
bosonic $Z_2$ twist fields, which are familiar from string
compactifications on orbifolds \orbf. For a single complex boson the
operator product expansion of the twist field $\sigma$ is 
\eqn\twist{\sigma(z) \partial X (z) = (z-w)^{-\half} \tau(w) +\cdots}
Here $\tau$ is an excited twist field. The conformal dimension of
$\sigma$ is $h=1/8$. The vertex operator for the  fermionic ground
state of the string stretched between the seven brane and the
D-instanton is then
given by
\eqn\vertdn{V_\chi= e^{-1/2\phi} \sigma_1\sigma_2\sigma_3\sigma_4
e^{-\half H_5}.}
Here $\sigma_i,i=1,\cdots,4$ are twist fields which change the
boundary conditions for the eight coordinates in the seven brane
directions  $X^{2i-2}+iX^{2i-1},\; i=1,\cdots,4$. The conformal dimension of
the vertex \vertdn\ is $h= 3/8+4\times 1/8+1/8=1$. Since a  stretched open
string has one end on the D7 branes and the other on  the D-instanton,
the Chan-Paton factors are transforming as $(8,  k)$ of
$SO(8)\times SO(k)$. The only nontrivial coupling of the vertices
\vertrep\fermcoll\ and \vertdn\ is given by the following three
point function on the disk.
\eqn\thrptf{\eqalign{\langle cV_\phi(x_1) cV\chi(x_2)
    cV_\chi(x_3)\rangle= \langle
ce^{-\phi} (\psi^8-i\psi^9)(x_1)\; c e^{-\half \phi} \prod_i \sigma_i
e^{i\half H_5}(x_2)\; ce^{-\half \phi} \prod_i \sigma_ie^{i\half
H_5}(x_3)\rangle}.}
Where $c$ denotes  the reparameterization ghosts, and inserting three
of them  fixes the Moebius
invariance of the disk amplitude.  The vertex $V^a$ in  \vertrep\ has been transformed into
the $-1$ picture in order to saturate the superghost anomaly on the
disk. To evaluate \thrptf\ note that the fermion in  can be bosonized
$\psi^8-i\psi^9= \exp(-i H_5)$. Using the free field correlators is is
easy to see that \thrptf\ is equal to a constant. Similarly the
calculations of three point functions reveals that two $\theta$
couple to $\phi_1+i\phi_2$ and two $\lambda$ couple to $\phi_1-i\phi_2$.

Denoting the wave functions of the bosonic  vertices $V^i$ and $V^a$ as $X^i$
and $\phi^a$, the fermionic vertices  $V^\alpha$ and  $V^{\dot
\alpha}$ with $\theta^\alpha$ and $\lambda^{\dot
\alpha}$ respectively, 
the matrix mechanics governing $k$ D-instantons near 4 D7 branes
on top of an orientifold seven plane is given by the following
action \eqn\matrac{\eqalign{S&= {1\over 2}\tr\Big(\half
[X_i,X_j]^2+[\phi_a,X_i]^2+\half[\phi_a,\phi_b]^2+ ig
\theta[(\phi_1+i\phi_2),\theta]\cr &+ig
\lambda[(\phi_1-i\phi_2),\lambda]+ig
\theta\Gamma_i[X^i,\lambda]\Big)+ ig \chi_I (\phi_1-i\phi_2)\chi_I.}} 
This matrix action can also be deduced by applying a T-duality on
action of the type I$^\prime$  D0 brane quantum mechanics \zeroeight\bgs.
The fields $X^i$ and $\theta^\alpha$ transform  in the symmetric
second rank tensor representation of $SO(k)$ have a trace part which
decouples from  \matrac\ and are   'center of mass'
degrees of freedom. In particular the trace part of  the fermion
$\theta^\alpha$ can be viewed as the fermionic collective
coordinates associated with the broken supersymmetries of the
D-instanton in the D7-O7 background. In the absence of a second
quantized formulation of string theory the treatment of the
D-instanton moduli space involves a certain amount of guesswork. Using
  ideas which have worked well for IIB instantons \gutc, we will
define an integration over the collective coordinates as an integral
over the matrix degrees of freedom associated with the D-instanton
weighted with the Matrix action \matrac.
 Note that the trace part of
$\theta$ does not appear in the matrix action \matrac\ hence the
integration over these fermionic variables has to be saturated by
additional insertions which will produce instanton induced interaction
vertices.

Another important feature is the coupling of the bosonic field
$\phi_1\pm i \phi_2$ to the fermnionic degrees of freedom
$\theta,\lambda$ and $\chi$. The fermionic integrals are saturated by
pulling bilinears in the fermions from the action. This gives only a
non vanishing result if there are an equal number of $\phi_1+ i \phi_2$
and  $\phi_1- i \phi_2$ appearing  in this process, since otherwise
the phase integrations of $\phi$ will kill the matrix integral.
Since for  $SO(2k)$ matrix mechanics the number of $\theta_\alpha$ is  $8\times 2k(2k+1)/2-8$
 and the number of  $\lambda$ is  $4\times 2k(2k-1)/2$ it follows that we
need to pull down $8\times (2k-1)$ $\chi$'s from the action \matrac. Note that
there are $8\times 2k$ $\chi$ altogether. This implies that the
integration over  eight $\chi$ cannot be saturated by the
action and has to come from other insertions. Note that the $\tr(
\theta\Gamma_i[X^i,\lambda])$ term in \matrac\ saturates an equal
number of $\theta$ and $\lambda$ integrations and does not change the 
counting of the deficit  above.

In summary  the integration over eight trace components of $\theta$ and
eight $\chi$ have to be saturated by additional insertions. These
insertion will be chosen 
here to produce the $\tr F^4$ and $(\tr F^2)^2$ instanton induced interactions.
  Since the $SO(8)$ gauge fields are associated with the seven
brane and the fermionic zero modes with the D-instanton, the simplest
string diagram which involves both is given by the disk with two
boundary changing (twist) operators \vertdn\ inserted such that a part of the
boundary of the world sheet ends  on the seven brane and a part ends on
the D-instanton. The simplest amplitude involves  
one gauge field vertex $V_{F}= F_{ij}(X^i\partial X^j+i \psi^i\psi^j)$
and two fermionic zero modes together with two 
boundary condition changing operators  \vertdn\ inserted. This
amplitude   has the form (only the $F_{ij}\psi^i\psi^j$ part of the
vertex contributes).
\bigskip
\ifig\fone{A disk with two twist operators, one seven brane gauge
  field and two fermionic zero modes inserted}
{\epsfbox{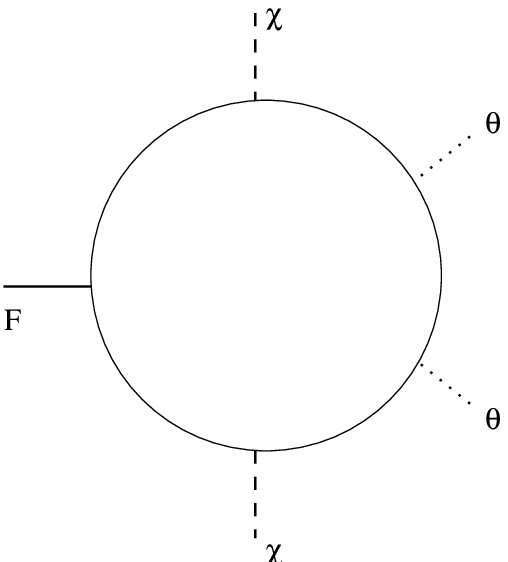}}
\eqn\fermzm{\eqalign{A&=F^A_{ij} T^A_{IJ} \theta^a\theta^b  \int dx_2 dx_4\langle c
  e^{-1/2\phi}S^a(x_1)\; e^{-1/2\phi}
V_\chi^I(x_2)\; c \psi^i\psi^j(x_3)\; e^{-1/2\phi}
V_\chi^I(x_4)\; c e^{-1/2\phi}S^b(x_5)\rangle\cr
&= \int_{x_1}^{x_3} dx_2 \int_{x_3}^{x_5} dx_4 {x_5-x_1\over (x_4-x_2)\sqrt{(x_2-x_1)(x_4-x_1)(x_5-x_2)(x_5-x_4)}} 
F_{ij}^A \chi^IT^A_{IJ}\chi^J \theta\gamma_{ij}\theta .}}
In evaluating \fermzm\ the Moebius invariance of the disk amplitude is
fixed by inserting three reparameterization ghosts $c$ which fix  the
positions of the vertices at $x_1,x_3$ and $x_5$. The correlators are
evaluated using
standard results for the   superghosts, spin fields and
twist fields  \friedan\orbf. Choosing the positions of
the fixed vertex operators to be $x_1=0,x_3=1$ and $x_5=\infty$ the
remaining integrations over $x_2$ and $x_4$ in \fermzm\ are
elementary and give 
\eqn\intdisk{\int_0^1 dx_2\int_1^\infty dx_4 {1\over (x_4-x_2)
    \sqrt{x_2x_4}}= \int_1^\infty {dx_4\over
      x_4}\ln\left({\sqrt{x_4}+1\over
        \sqrt{x_4}-1}\right)={\pi^2\over 2}.}
The eight fermionic zero modes $\theta^a$ associated with
the supersymmetries  broken by the presence of the D
instanton are soaked up by four disk diagrams \fermzm. Furthermore  the four
disk diagrams also  saturate the integrations over the eight remaining
$\chi$  and  induce 
a term with four $F$ gauge fields. 
Note that the integral over $\theta^\alpha$ reproduces the well known
kinematic tensor 
$t_8$ since
\eqn\invtens{\int d^8\theta\; \theta\gamma^{ij}\theta\;
  \theta\gamma^{kl}\theta \; \theta\gamma^{mn}\theta \;
  \theta\gamma^{pq}\theta = t_8^{ijklmnpq}.}
In the simplest case of  of $SO(2)$ matrix mechanics it is
straightforward to check that the integration over $\chi$ produces the
factors $\tr(F^4)$ and $(\tr F^2)^2$ in the correct relative
normalization as determined by \seveninst.  It is a natural
generalization of the arguments involving type IIB thresholds and
$SU(N)$ matrix integrals \gutc\  that the threshold
amplitudes \seveninst\ are related to matrix integrals with action
\matrac. Since the matrix mechanics for $SO(2k)$ for
general $k$ is very complicated we do not attempt to check this
correspondence here, although it would be very interesting to try to
apply the methods of \nikita\ to evaluate the matrix integrals for
arbitrary $k$. Note that because of the insertions described above,
 a certain correlation function involving eight
$\chi$ and not the
partition function itself has to be compared to the results given  in
this paper. 
\newsec{D-Instanton in supergravity}
The D-instanton
solution of IIB  supergravity \pgg\ is  a BPS solution which preserves
half the  supersymmetries,
\eqn\sugreq{\delta \lambda =iP_\mu \epsilon^*,\quad \delta\psi_\mu = D_\mu
\epsilon.}
Where the scalar field strength, covariant derivative and composite
$U(1)$ gauge field are defined as
\eqn\notst{P_\mu = i {\partial_\mu \tau\over 2\tau_2},\quad D_\mu \epsilon
= (\partial_\mu +{1\over 4} \omega_\mu^{ab} \Gamma_{ab} -{1\over 2}i
Q_\mu)\epsilon,\quad Q_\mu = -{\partial_\mu \tau_1\over 2\tau_2}, }
and the complex scalar given by a combination of RR scalar  and dilaton
$\tau = \chi+i e^{-\phi}$. 
Solutions to \sugreq\ which are BPS satisfy (in Euclidean space)
$\partial_\mu  \chi= \pm i \partial_\mu e^{-\phi}$. The D-instanton
  solution  is then 
$g_{\mu\nu}=\eta_{\mu\nu}$ and the dilaton satisfies 
\eqn\disnteqa{\partial_\mu\partial^\mu e^{\phi}=0 }
i.e. the dilaton profile for a D-instanton centered at $y^\mu$  is given in terms of a harmonic function
\eqn\disnteqb{e^{\phi}(x)= e^{\phi_0}+{2K\alpha^{\prime 4}\over \pi^4 |x-y|^8}.} 

 When the three other singularities are far away the geometry
 in the vicinity of one of the  orientifold singularities
 on   $T^2/Z_2$  is locally
  $R^2/Z_2$. In the
supergravity this implies that under the inversion $I$  $x_a\to -x_a,a=8,9$
transform according to the orientifold projection $(-1)^{F_L}\Omega
I$. In particular this relates at $z$ to fields at $-z$. From the  32
supersymmetries    of IIB   only sixteen satisfying 
\eqn\susyorb{\eta = (1+i \Gamma^{89})\epsilon,}
are unbroken by the orientifold projection.

Since the dilaton and RR scalar sources of of four D7 branes on top of
the orientifold cancel locally, the D-instanton solution of IIB  \disnteqb\ can be
used to write down a  solution of the orientifolded IIB supergravity.

We have to find a solution which is invariant under the orientifold
 projection acting on the fields of IIB supergravity. 
 Since there is no nontrivial monodromy
of $\tau$ under the orientifold projection it relating  the
 fields at $z$ and $-z$ . Therefore the simplest invariant  dilaton profile 
\eqn\disnteqc{e^{\phi}(x^i,z)= e^{\phi_0}+{2K\alpha^{\prime 4}\over
\pi^4 ((x-x_1)^2+|z-z_1|^2)^4}+{2K\alpha^{\prime 4}\over
 \pi^4((x-x_1)^2+|z+z_1|^2)^4}.}
Where we split the coordinate $x^\mu,\mu=0,\cdots,9$ into
 $x^i,i=0,\cdots,7$ and $z,\bar{z}$.
It is obvious that\disnteqc\ is  invariant under $z\to -z$. Since the solution
 \disnteqb\ satisfies the charge quantization condition minimally,
 i.e. corresponds to singly charged D-instanton, it is easy to see
 that the projected  instanton
solution  \disnteqc\ has instanton charge 2. A 'stuck' instanton can
carry instanton charge one 
\eqn\disnteqc{e^{\phi}(x,z)= e^{\phi_0}+{2K\alpha^{\prime 4}\over \pi^4
((x-x_1)^2+|z|^2)^4},}
and 
is also invariant under $z\to -z$ but cannot move away from the
orientifold singularity $z=0$.

In \bianchi\ the D-instanton solution in $AdS_5\times S_5$ was
discussed. Since $AdS_5\times S_5$  is conformally flat the solution
can easily be obtained from the flat space D-instanton solution. Here
we shall only be interested in the appropriately rescaled D-instanton
solution in the limit when the profile is evaluated at the boundary of $AdS_5$.
\eqn\disntads{\lim_{\rho\to 0} \rho^{-4}\big( e^\phi-e^{\phi_\infty}\big)=
  {2K \alpha^{\prime 4}\over L^8 \pi^4} {\rho_0^4\over
    ((x-x_0)^2+\rho_0)^4}.}
Here $\rho_o,x_0$ label the position of the D-instanton in the bulk
and $x$ is a coordinate on the boundary of $AdS_5$. Note that in the
limit $\rho\to 0$ the position on the $S^5$ has disappeared from
\disntads. In particular this implies that even in the orientifolded
$AdS_5\times S^5$ the D-instanton solution will be of the form
\disntads\ (with twice the charge). 
\newsec{$N=2$ $USp(N_c)$ theories and the AdS/CFT correspondence}
As mentioned in the introduction  a probe of $N_c$  D3 brane
 moving in 
 Sen's $Z_2$ IIB orientifold produces  a $USp(2N_C)$ $N=2$  gauge theory on the
worldvolume  with four
hypermultiplets transforming in the fundamental and one hypermultiplet
transforming in the second rank antisymmetric tensor of  $USp(2N_c)$
\probe.  The $SO(8)$ gauge symmetry on the D7 branes corresponds to a
global SO(8) symmetry of the four fundamental hypermultiplets.

The $USp(2N_c)$ $N=2$ with the matter content described above has vanishing beta
function and for zero Higgs expectation values defines a conformal
field theory for any integer value of  $N_c$. Hence we can consider the large $N$
limit. In \spal\aharony\  the AdS/CFT correspondence was generalized
to this conformal 
field theory and it was shown that the  $USp(2N_c)$ $N=2$ theory is dual
to IIB supergravity on $AdS_5\times S_5/Z_2$. The metric on
the $S_5/Z_2$ is given by
\eqn\metricsphere{ds^2= d\theta^2 +\sin^2(\theta) d\phi^2 +cos^2(\theta)d\Omega_3^2,}
here $\phi\in[0,\pi]$ is periodic with  period $\pi$ instead of $2\pi$ for an
ordinary $S_5$ and some of the fields of IIB
supergravity have nontrivial monodromies as $\phi\to \phi+\pi$ \aharony.
 The $S^3$ can be regarded as
the fixed  point of the orientifold and the original sevenbranes are
filling $AdS_5$ and are wrapped on the $S^3$ at $\theta=0$. The
changed 
periodicity and monodromy of type IIB fields  modifies  the spectrum
of the chiral primaries 
coming from the bulk of the supergravity as analyzed in
\spal\aharony. In addition there are  chiral primaries coming from fields
localized on the seven 
branes \aharony. Such states carry $SO(8)$ charges. All fields in the
supergravity and field theory are characterized by the charges they
carry with respect to 
\eqn\symmr{SU(2)_R\times SU(2)_L\times
U(1)_R\times SO(8)\times USp(2N_c),}
where $SU(2)_R\times U(1)_R$ is identified with the R-symmetry of the
the $N=2$ superconformal algebra  and $SU(2)_L$ can be 
interpreted as a flavor symmetry of the AST hypermultiplet, $SO(8)$
is a gauge symmetry on the D7 branes which is a global flavor
symmetry from the perspective of the D3 branes and 
 $USp(2N)$ denotes the gauge symmetry of the D3 brane. Geometrically
 the $SU(2)_L\times SU(2)_R$ can be  
identified with the $SO(4)$ symmetry of the $S^3$.

On the AdS side we  will mainly be interested in the fields related to
the vector field $A_M$ of the heterotic string. In the $AdS$ limit this
vector field decomposes into three 
different fields which are all transforming in the adjoint of
$SO(8)$,

Firstly the $M=0,\cdots,4$ components of the vector state are related to
a vector field in $AdS_5$ with  $A_\mu=\sum_k
A_\mu^k(x) Y^k(y) $ which  transforms as $(k,k)_0$ under  under
$SU(2)_R\times SU(2)_L\times 
U(1)_R$. Where $Y^k(y)$ is  the  $k$-th  scalar KK mode on the 3
sphere.  
The $M=5,6,7$
components are related to  vectors with   polarizations taking values
in the tangent space of  $S_3$, These states lead to  scalar fields in the
$AdS_5$. The  $A_a= \sum_k A_k(x) Y^k_a(y)$ transform  as
$(k,k+2)_0+(k+2,k)_0$. Where $Y^k_a$ is the $k$-th vector spherical
harmonics on $S^3$. 
$M=8,9$ components are related to the scalar $z= \sum_k z_k(x)Y^k(y)$ 
which transforms as $(k,k)_2$.
 
The part of the $k=1$ KK mode of the internal vector $A_a$ which
transforms as $(3,1)_0$ represents a chiral primary with dimension
$\Delta=2$ \foot{A WZ coupling of the RR four form potential and the
gauge field in the seven brane  leads to a split of the $(3,1)_0$ and $(1,3)_0$ and a
shift of their masses, and only the $(3,1)_0$ component corresponds to
a chiral primary with $\Delta=2$ \aharony.} In the gauge theory side  these fields
are identified with composite 
operators which are chiral primaries and have the same dimension and quantum
numbers.  $A_a^{k=1}$  is associated with the bilinear made of $q^I$
\eqn\zdefin{z^{[IJ]}= q^{[I}_Aq^{J]}_BJ^{AB}.}
Where $q^I$ are the scalars in the fundamental hypermultiplets, and
$J^{AB}$ is the invariant second rank $USp(2N)$ tensor and we have
suppressed the $SU(2)_R$ indices. 
Other members of this supermultiplet can be obtained by acting with
supercharges on the chiral primary field \zdefin.  Of  particular interest  the $k=1$ vector field $A_\mu^{k=1}$ transforms in the $(1,1)_0$ and has
conformal dimension $\Delta=3$. 
The vector field $A_m^{k=1}$  is identified with the global SO(8) current
\eqn\current{J_\mu^{[IJ]} = q^{[I}\partial_\mu q^{J]}+ i \psi^{[I}\gamma_\mu
  \psi^{J]},}
which is given by acting with  $Q\bar{Q}$ on the chiral primary
\zdefin. 
Furthermore the $z^{k=1}$ scalar is associated with fields $qXq$ and
$\psi\psi$ and is given by acting with  $Q^2$ (or $\bar Q^2$ for the conjugate) on \zdefin.

The  four derivative threshold corrections in the heterotic
string \seveninst\ are mapped to  D-instanton induced  $F^4$ terms
living on the sevenbrane in the IIB orientifold. 
In the AdS/CFT correspondence the seven branes wrap the $S^3$ and fill
$AdS_5$ and they lead to new bulk four point vertices in the
$AdS_5$. Using the relation of the heterotic vector fields to the 
the vector $A_m$ and and the scalars $A_a$ and $z$, \seveninst\ will
lead to four point correlation functions of these fields. 

In the following the four point function involving four vector fields
$A_m$ in $AdS_5$ are considered in detail, the four point vertices are
given be the following expression, 
\eqn\vecads{I_4= \int {d^4z dz_0\over z_0^5}z_0^8
t_8^{mnpqlrst}D_mA^A_nD_pA^B_qD_lA^C_r D_sA^D_t \Big(F_{ABCD}(\tau)+
G_{ABCD}(\tau)\Big).}
Here the indices $m,n,\cdots=0,\cdots 4$ denote coordinates in $AdS_5$
and the superscript in $A^A_n$ labels a basis of the adjoint
representation of $SO(8)$. The tensor $t_8$ is constructed using
$\delta_{mn}$ and the  factor $z_0^8$ in \vecads\ comes from four
inverse metrics. The functions  $F_{ABCD}$ and $G_{ABCD}$
in \vecads\ contain the information about the D-instanton contribution
and have the following
structure 
\eqn\instacfg{\eqalign{F_{ABCD}(\tau)&= \tr(t_A t_B
    t_Ct_D)\sum_N\Big({1\over 2}\sum_{N|m}{1\over m} e^{2\pi i 
2N\tau}-{1\over 2}\sum_{N|m}{1\over m} e^{2\pi i 4N\tau}+c.c.\Big)\cr 
G_{ABCD}(\tau)&= \tr(t_At_B)\tr(t_Ct_D) \sum_N\Big({1\over
  4}\sum_{N|m}{1\over m} e^{2\pi i 
2N\tau }-{1\over 8}\sum_{N|m}{1\over m} e^{2\pi i 4N\tau}+c.c.\Big).}}
Using the prescription developed in \adscft\ the vertices \vecads\
contribute to  a correlation function in the CFT of four $SO(8)$
currents, $\langle J^{\mu A}(x_1)J^{\nu B}(x_2)J^{ \rho C}(x_3)J^
{ \lambda D}(x_4)\rangle$.
This calculation  uses  the bulk to boundary propagator for a vector field in
$AdS_5$, given by $G_{m\mu}(z,x)$ which relates the bulk gauge field
$A_m(z)$ to the boundary values $A_\mu(x)$. The bulk to boundary
propagator defined  in \freedman\ has the following form 
\eqn\fregr{G_{m\mu}(z,x)= {3\over \pi^2} {z_0^2\over
    \big(z_0^2+(z-x)^2\big)^3} J_{m\mu}(z-x).}
Note that in \vecads\ only the field strength $D_{[m}G_{n]}\mu(z,x)$
  appears  and this expression is  independent of the choice of gauge
  in \fregr. In addition  the covariant derivative can be replaced by
an ordinary derivative and the result can be expressed in a simple
manner \fregr.
\eqn\frefi{D_{[m}G_{n]\mu}(z,x)= \partial_{[m}G_{n]\mu}(z,x)={6\over \pi^2}{z_0\over
    \big(z_0^2+(z-x)^2\big)^3}J_{o[m}(z-x) J_{n]\mu}(z-x).}
The conformal tensor $J_{m\mu}$ is defined by
\eqn\jmmudef{J_{m\mu}=\big(z_0^2+(z-x)^2\big) {\partial\over \partial
    z_m} \left( {(z-x)_\mu\over \big(z_0^2+(z-x)^2\big)}\right), }
and $(z-x)_0=z_0$ is implied in \jmmudef.
Plugging \frefi\ into \vecads\  gives the following contribution to the
four point function of  $SO(8)$ currents
\eqn\fouptsoe{\eqalign{&\langle
    J_A^\mu(x_1)J_B^\nu(x_2)J_C^\rho(x_3)J_D^\lambda(x_4)\rangle=
    t_8^{mnpqlrst} \left({6\over \pi^2}\right)^4\int {d^4z dz_0\over
      z_0^5}   z_0^{12}    {J_{o[m}(z-x_1) J_{n]\mu}(z-x_1)\over
    \big(z_0^2+(z-x_1)^2\big)^3}\cr
&\times {J_{o[p}(z-x_2) J_{q]\nu}(z-x_2)\over
    \big(z_0^2+(z-x_2)^2\big)^3}
{J_{o[l}(z-x_3) J_{r]\rho}(z-x_3)\over
    \big(z_0^2+(z-x_3)^2\big)^3}{J_{o[s}(z-x_4) J_{t]\lambda}(z-x_4)\over
    \big(z_0^2+(z-x_4)^2\big)^3}\cr
&\times \Big\{ F_{ABCD}(\tau)+ G_{ABCD}(\tau)\Big\}. }}
This constitutes a prediction for the instanton contributions to the
four point correlator of four $J^{\mu A}$. It is easy to generalize
this calculation to the four point functions involving the internal
vector $A_a$ and the scalar $z$ using \seveninst, as well as all
Kaluza-Klein descendants.

\newsec{D-instantons and Yang-Mills instantons}
The map between the parameters of SYM and IIB string theory on
$AdS_5\times S_5$ is 
given by
\eqn\mappara{g_a = q_{YM}^2/4\pi,\quad \chi = \theta_{YM}/2\pi,\quad
  R^2/\alpha^\prime = \sqrt{q_{YM}^2 N}.}
A first indication that D-instanton effects are related to
YM-instanton effects is given by the observation  that   the action for a charge k D-instanton  $\exp(-2\pi k
/g_s)$ is mapped to the charge k YM-instanton action $\exp(-8\pi k
/g_{YM}^2)$ \banksgreen. In \bianchi\ more evidence for this  correspondence
 was found  by considering the
D-instanton solution of IIB supergravity in the $AdS_5\times S_5$
background. 
The gauge field $\tr(F^+)^2$ of an $SU(2)$ instanton of charge one is given by 
\eqn\profile{\tr(F^-)^2= {4\over g^2_{YM}} {\rho_0^4\over
    (\rho_0^2+(x-x_0)^2)^4}.}
Remarkably agrees with the D-instanton solution in $AdS_5\times S^5$ \disntads, where $x$ a point at the
boundary and $(\rho_0,x_0)$ is the location of the D-instanton in the
bulk of $AdS_5$. Note that the scale size $\rho$ of the YM instanton
is  the position of D-instanton in the radial direction of
$AdS_5$. This relation is one example of the IR/UV relation in the 
gauge theory/supergravity correspondence. 

In an impressive  series of papers \mattis\ the predictions for instanton
contributions to correlators in the large $N_c$ limit of $N=4$
$SU(N_c)$ SYM were 
confirmed using ADHM multi instanton calculus.
One important fact  which makes this correspondence possible is that  the
threshold corrections like $t_8t_8R^4$ in IIB only receive
perturbative contribution at tree level and one loop. This implies
that one can isolate the instanton contributions reliably even in the
large $N_c$ limit.

For the $N=2$ conformal field theory which arises in the IIB
orientifold discussed in this paper a similar
calculation of the instanton contribution in the gauge theory should
be possible. Note that for $USp(2N_c)$ theories with hypermultiplets
transforming in the fundamental representation there is a flavor
parity symmetry \seibwit\mattisb\  which implies that only  even instanton 
numbers contribute to these instanton induced interactions, in agreement with  \seveninst.

A detailed analysis of the multi instanton ADHM construction for
$USp(2N_c)$ $N=2$ SYM with the matter content described above is beyond
the scope of the present paper. It would be nevertheless very
interesting to see whether all or some of the features of the
impressive $N=4$
analysis \mattis\ generalize  in the case at hand. There are eight (instead
of sixteen) zero modes associated with the broken superconformal
symmetries which when soaked up by operator insertions should produce
the four point correlators discussed above. All the other fermionic
zero modes have to be lifted by quadrilinear terms in the multi
instanton action.  It would be interesting to see whether in the
large $N_c$ saddle point approximation the dominant contribution comes
from a single copy of $AdS_5$ and whether 
integrals in the $k$ instanton partition functions appear producing  the
$SO(2k)$ matrix integral \matrac.

\newsec{Conclusions}
In this paper D-instanton induced terms in the worldvolume of D7
branes in IIB orientifold were obtained by heterotic/type I duality
from a one loop heterotic calculation. In the following we attempted
to apply the methods
which were applied successfully for thresholds in IIB and their
relation to D-instantons as well as the relation of D-instantons in
$AdS_5\times S_5$ and YM-instantons in $N=4$ SYM. It is not clear
whether this approach is justified since no independent check of the
prediction coming from the heterotic loop calculation have been
performed, unlike in the case of $N=4$ SYM. It would therefore be very
interesting to perform such (rather nontrivial) checks.
It would also be interesting to analyze the type I formulation of the
theory from the matrix string perspective \pierre\gava.

The arguments  presented in this paper can be generalized in various
ways. It is easy to calculate thresholds involving four  gravitons,
here the situation is more involved since there can also be
contributions from the ten dimensional bulk as well as higher
curvature terms living on the three branes \bbg.
The orientifold theory presented here is rather special it is an open
question whether other heterotic loop calculations  can be related to
interesting quantities interesting
conformal theories with $N=2$ or $N=1$ susy by similar 
arguments given in the paper.
\medskip
\noindent{\bf Acknowledgments}
\medskip
This work was supported in part by NSF grant PHY-9802484. 
\medskip

\appendix{A}{Details of the heterotic loop amplitude}
In this appendix  the calculation of one loop heterotic amplitudes
  with four gauge fields is outlined. The method is adapted from \gutb\ for the Wilson line
  defined \wilsonl\ and differs only slightly from the method used in
  \lerche. The reader who is only interested in the result
  \seveninst\ may want to skip the details given in the appendix.
\medskip
  The integrals that will appear in this
calculations are of the following form
\eqn\oneloopfour{I_{Q}=\int_F{d^2\tau\over
      \tau_2} \sum_{A} {T_2\over \tau_2} \exp\Big\{ 2\pi i T
      \det{A}-{\pi T_2\over \tau_2 U_2}\big| (1 U) A
          \Big(\matrix{\tau \cr 1}\Big)\big|^2\Big\}  Q C(Y,A).}
Here the matrix A is given by  2$\times$2 matrices with integer
entries
\eqn\amat{A= \pmatrix{m_1&  n_1\cr m_2 &n_2},\quad m_1,m_2,n_1,n_2\in Z,}
 and $C(Y,A)$ is the partition function of the $SO(32)$ lattice which
  in general depends on the Wilson line $Y$ defined in \wilsonl\ and the matrix
 $A$ \amat\ in the following way
\eqn\intlat{\eqalign{C(Y,A)&=  \sum_{a,b=0,1}\prod_{k=1}^{16}
e^{-i\pi(m^in^jY_i^k Y_j^k+b n^iY_i^k)} \theta\left[\matrix{a+2m^l
Y_l^k\cr b+2n^l Y_l^k}\right](0,\tau) \cr &=
\sum_{a,b}\theta^4\left[\matrix{a\cr
b}\right](0,\tau)\theta^4\left[\matrix{a+m_2\cr
b+n_2}\right](0,\tau) \theta^4\left[\matrix{a+m_1\cr
b+n_1}\right](0,\tau)\theta^4\left[\matrix{a+m_1+m_2\cr
b+n_1+n_2}\right](0,\tau).}}  Where
the standard notation for the theta functions is introduced
\eqn\thetnot{\theta\left[\matrix{1\cr
      1}\right]=\theta_1,\quad\theta\left[\matrix{1\cr
      0}\right]=\theta_2,\quad \theta\left[\matrix{0\cr
      0}\right]=\theta_3,\quad\theta\left[\matrix{0\cr
      1}\right]=\theta_4,\quad.}
The form of the operator $Q$ in \oneloopfour\ depends on the
threshold in question. The operator $Q$ for $\tr(F^4)$ and
$(\tr(F^2))^2$ can be found by 'gauging' \intlat \ellis. The
Wilson line \wilsonl\ breaks the gauge group to $SO(8)^4$ and the
thirty two free fermions of the $SO(32)$ lattice are split into
four sets of eight in \intlat. The result depends on the spin
structures $[a,b]$ for the eight fermions which are associated
with the first $SO(8)$ in \intlat. For the $\tr(F^4)$ threshold
the operators are given by
\eqn\qresult{\eqalign{Q_{\tr(F^4)}\left[\matrix{1\cr
        0}\right](\tau)&=-{1\over 2^8 3}
  \theta_3^4\theta_4^4(\tau),\cr
 Q_{\tr(F^4)}\left[\matrix{0\cr 0}\right](\tau)&={1\over
     2^8 3} \theta_2^4\theta_4^4(\tau),\cr
Q_{\tr(F^4)}\left[\matrix{0\cr   1}\right](\tau)&=-{1\over   2^8 3}
\theta_2^4\theta_3^4(\tau), }}
whereas for the $(\tr(F^2))^2$ threshold the operator is given by
\eqn\trftwoq{\eqalign{Q_{(\tr(F^2))^2}\left[\matrix{1\cr
0}\right](\tau)&={1\over 2^{10}3^2}\big(e_2(\tau)+\hat{E}_2(\tau)\big)^2,\cr
 Q_{(\tr(F^2))^2}\left[\matrix{0\cr
0}\right](\tau)&={1\over 2^{10}3^2}\big(e_3(\tau)+\hat{E}_2(\tau)\big)^2,\cr
Q_{(\tr(F^2))^2}\left[\matrix{0\cr
1}\right](\tau)&={1\over 2^{10}3^2}\big(e_4(\tau)+\hat{E}_2(\tau)\big)^2.}}
 Where the following
notation has been introduced \eqn\edefin{e_2=
\theta_3^4+\theta_4^4,\quad e_3= \theta_2^4-\theta_4^4,\quad e_4=
-\theta_2^4-\theta_3^4,}
and $\hat{E}_2$ is the nonhomolomorphic (but modular) Eisenstein
function of weight two defined as $\hat{E}_2= {E}_2- 3/(\pi \tau_2)$.
In the following it will be useful to expand \trftwoq\ in powers of
$1/\tau_2$.
\eqn\exptrft{\eqalign{Q_{(\tr(F^2))^2}\left[\matrix{a\cr
b}\right](\tau)= \sum_{r=0,1,2} {1\over \tau_2^r}{Q^{(r)}_{(\tr(F^2))^2}\left[\matrix{a\cr
b}\right](\tau) }}}
For example it is easy to see that $Q^{(0)}_{(\tr(F^2))^2}$ is given by \trftwoq\
  with $\hat{E}_2$ replaced by $E_2$ and 
\eqn\exptrftb{Q^{(1)}_{(\tr(F^2))^2}\left[\matrix{1\cr
0}\right](\tau)= -{1\over 2^9 3
\pi}\big(e_2(\tau)+E_2(\tau)\big),\quad Q^{(2)}_{(\tr(F^2))^2}\left[\matrix{1\cr
0}\right](\tau)= -{1\over 2^{10}\pi^2}}
and similarly for the other charge insertions in \trftwoq.
\appendix{B}{Evaluation of integral}
The integral \oneloopfour\ can be evaluated  using the method
of orbits \dixon. In the present context this technique was
discussed in \costas\kiritsis\ and in \lerche, where type I
thresholds  with certain Wilson lines present were evaluated using
results from \stephanb.  Without
Wilson lines it is straightforward to show that
 under the modular $SL(2,Z)$ transformations $\tilde{\tau}=
 (a\tau+b)/(c\tau+d)$ with $a,b,c,d\in Z$,$ad-bc=1$
   \eqn\stmat{{1\over \tilde{\tau}_2 }\big| (1 U) { A}
          \pmatrix{\tilde{\tau} \cr 1}\big|^2= {1\over {\tau}_2 }\big|
          (1 U)  {A} \pmatrix{a&b\cr c&d}
          \pmatrix{{\tau} \cr 1}\big|^2.}
 The summation over all integer matrices  matrices ${ A}$ can then
 replaced  by the summation over all equivalence classes of
 $SL(2,Z)$ orbits.  There are
three different cases, the trivial orbit ${A}=0$, the degenerate
orbit $\det({A})=0$ and the non degenerate orbit $\det({A})\neq 0$

In the following we will consider only  the non degenerate orbit, where
the  fundamental ${\cal F}$ is unfolded into the double cover of the
upper half plane ${\cal H}$. The non degenerate $SL(2,Z)$ orbits fall
into the following equivalence classes
\eqn\orbitrf{{ A}=\pm \pmatrix{k&j\cr 0& p},\quad k>0, 0\leq j<k, p\in Z.}
When Wilson lines are present, matters are more complicated but using
the well known transformation properties of the theta functions under
$\tau\to \tau+1,\tau\to -1/\tau$ is is easy to see that for both
$Q_{\tr(F)^4}$ \qresult\ and $Q_{(\tr(F)^2)^2}$ \trftwoq, $QC(Y,A)$
defined in \intlat\ behaves in the following way
\eqn\trafq{QC(Y,{A})({a\tau+b\over c\tau+d})=QC(Y,{ A}
  \pmatrix{a&b\cr c&d} )(\tau).}
Hence the method of orbits can be used to unfold the integral. For
the non degenerate orbit we get
\eqn\unfoldedint{I_{nd}=\int_{H}{d^2\tau\over
      \tau_2} \sum_{\quad k>0, 0\leq j<k, p\in Z}{T_2\over \tau_2}
       \exp\Big\{ 2\pi i kpT
      -{\pi T_2\over \tau_2 U_2}\big| k\tau +j +pU|^2\Big\} Q
      C(Y,\pmatrix{k&j\cr0&p})(\tau).}
In order to evaluate \unfoldedint\ it is convenient  to  split the summation over
equivalence classes ${ A}$ in \orbitrf\ into four separate sectors  ${
  A}^{(i)},i=1,\cdots,4$.
\eqn\amas{\eqalign{A^{(1)}&= \pmatrix{2\tilde{k} & 2\tilde{j} \cr 0
      &2p},\quad\quad 0\leq \tilde{j}<\tilde{k},\cr
A^{(2)}&=\pmatrix{2\tilde{k}+1&  2\tilde{j}\cr 0 &2p},\quad 0\leq
 \tilde{j}\leq\tilde{k},\cr
 A^{(3)}&=\pmatrix{2\tilde{k}+1&
    2\tilde{j}+1\cr 0 & 2p },\quad 0\leq \tilde{j}<\tilde{k},\cr
\quad A^{(4)}&= \pmatrix{2\tilde{k} & 2\tilde{j} \cr 0
      &2p+1},\pmatrix{2\tilde{k}&  2\tilde{j}+1\cr 0
    &2p+1},\pmatrix{2\tilde{k}&  2\tilde{j}+1\cr 0
    &2p}\quad 0\leq \tilde{j}<\tilde{k}.}}
\medskip
\noindent{\bf B1. The $\tr(F^4)$ threshold}
\medskip
 Using \intlat\ and \qresult\ we can
express $QC(A^{(i)})$ for the $\tr(F_1)^4$ threshold as
\eqn\charge{\eqalign{QC_1=Q C(A^{(1)})&= {1\over  2^8 3}{1\over
 \eta^{24}}\big(-\theta_2^{16} 
  \theta_3^4\theta_4^4 +\theta_3^{16}
  \theta_2^4\theta_4^4- \theta_4^{16}
  \theta_2^4\theta_3^4\big)= 1,  \cr
QC_2=Q  C(A^{(2)})&= {1\over  2^8 3}{1\over  \eta^{24}}\theta_2^8 \theta_3^8(
  -\theta_3^4\theta_4^4 +\theta_2^4\theta_4^4)
  = -{1\over 3}, \cr
QC_3=Q  C(A^{(3)})&={1\over 2^8 3}{1\over \eta^{24}}\theta_2^8 \theta_4^8(
  -\theta_3^4\theta_4^4 -\theta_2^4\theta_3^4)
  = -{1\over 3}, \cr
QC_4=Q  C(A^{(4)})&={1\over  2^8 3 }{1\over \eta^{24}}\theta_3^8 \theta_4^8(
  \theta_2^4\theta_4^4 -\theta_2^4\theta_3^4)
  =- {1\over 3}.}}
Where  the following  identities were used 
\eqn\identone{\theta_2^4+\theta_4^4-\theta_3^4=0,\quad
 \theta_2^4\theta_3^4\theta_4^4=16\eta^{12}, \quad 
  \theta_3^{12}-\theta_2^{12}-\theta_4^{12}=48\eta^{12}.}
Using the (trivial) fact that $Q  C_1=Q  C_2+ Q  C_3+Q
C_4+2$. The summation can be rearranged such that $k,j$ and $p$ run
 over the original summation range defined in \orbitrf.
\eqn\unfoldedinb{\eqalign{I^{\tr(F^4)}_{nd}&=2\int_{H}{d^2\tau\over
      \tau_2} \sum_{\quad k>0, 0\leq j<k, p\in Z}{T_2\over \tau_2}
       \exp\Big\{ 2\pi i 4 kpT
      -{\pi4 T_2\over \tau_2 U_2}\big| k\tau +j +pU|^2\Big\}\cr
&-\int_{H}{d^2\tau\over
      \tau_2} \sum_{\quad k>0, 0\leq j<k, p\in Z}{T_2\over \tau_2}
       \exp\Big\{ 2\pi i 2 kpT
      -{\pi2 T_2\over \tau_2 U_2}\big| k\tau +j +pU|^2\Big\} \cr
&=\sum_{N}\Big({1\over 2}\sum_{N|m} {1\over m}e^{-2\pi N 4T}- {1\over
 2}\sum_{N|m}{1\over m} e^{-2\pi N 2T}\Big)+cc }}
where we used the formula for $I_m$ derived in Appendix C for $m=4$
 and $m=2$ to evaluate the terms in  the first and second line of
 \unfoldedinb\ respectively.
\medskip
\noindent{\bf B2. The $(\tr(F^2))^2$ threshold}
\medskip
The operator $Q$ for  the $(\tr(F^2)_1)^2$ threshold
depending  on the spin structures was defined in  \trftwoq. Together
with \intlat\ $QC(A^{(i)})$ become
\eqn\ctrftwo{\eqalign{QC_1=Q C(A^{(1)})&=
{1\over 2^{10}3^2}{1\over
\eta^{24}}\Big\{\theta_2^{16}\big(e_2+\hat{E}_2\big)^2
  +\theta_3^{16}\big(e_3+\hat{E}_2\big)^2
  +\theta_4^{16}\big(e_4+\hat{E}_2\big)^2\Big\}, \cr
QC_2=Q  C(A^{(2)})&={1\over 2^{10}3^2} {1\over \eta^{24}}\theta_2^8
\theta_3^8\Big\{\big(e_2+\hat{E}_2\big)^2+\big(e_3
+\hat{E}_2\big)^2\Big\},
   \cr
QC_3=  QC(A^{(3)})&={1\over 2^{10}3^2}{1\over \eta^{24}}\theta_2^8
\theta_4^8\Big\{\big(e_2+\hat{E}_2\big)^2+\big(e_4+
\hat{E}_2\big)^2\Big\},\cr 
QC_4=Q  C(A^{(4)})&=   {1\over 2^{10}3^2}{1\over \eta^{24}}\theta_3^8
\theta_4^8\Big\{\big(e_3+\hat{E}_2\big)^2+\big(e_4+
\hat{E}_2\big)^2\Big\}.}}
Using the identity (see also \lerche)
\eqn\idenone{Q C_1=Q C_2+Q C_3 +QC_4-{1\over 2}} 
we can redistribute the summation over $A^{(1)}$ over the
contributions of
  the $A^{(i)},i=2,3,4$. The integral involving the constant term  on the right hand side of
  \idenone\ is then easily evaluated.
\eqn\unfoldedinc{\eqalign{I^{1}_{nd}&=-{1\over 2}\int_{H}{d^2\tau\over
      \tau_2} \sum_{\quad k>0, 0\leq j<k, p\in Z}{T_2\over \tau_2}
       \exp\Big\{ 2\pi i 4 kpT
      -{\pi 4T_2\over \tau_2 U_2}\big| k\tau +j +pU\big|^2\Big\} \cr
&= -{1\over 8}\sum_{N}\sum_{N|m}{1\over m} e^{-2\pi N 4T} }}
Where $I_m$ for $m=4$ form Appendix C was used.

To calculate the contribution to the integral of the terms containing
$QC_i,i=2,3,4$ we first   use  $Q
C_2(\tau-1)=Q C_3(\tau)$ to elimiate $Q C_3$ in favour of $QC_2$ with
an enlarged summation range. Keeping carefully track of  the range of
summations $A_i$ involved gives 

\eqn\oneloopfourb{\eqalign{I^{2+3+4}_{nd}&=\int_H{d^2\tau\over
      \tau_2} \sum_{A} {T_2\over \tau_2} \exp(2\pi i 2k T) 
      \Big\{ \exp\Big(-{\pi T_2\over \tau_2 U_2}\big| (1 U)
      \pmatrix{k&2j\cr 0&2p}
          \Big(\matrix{\tau \cr 1}\Big)\big|^2\Big)  Q C_2(\tau)\cr
&+ \exp\Big(-{\pi T_2\over \tau_2 U_2}\big| (1 U)
      \pmatrix{2k&j\cr 0&p}
          \Big(\matrix{\tau \cr 1}\Big)\big|^2\Big)  Q C_4(\tau)\cr
&+\exp\Big(-{\pi T_2\over \tau_2 U_2}\big| (1 U)
      \pmatrix{2k&j+k\cr 0&p}
          \Big(\matrix{\tau \cr 1}\Big)\big|^2\Big)  Q C_4(\tau)\Big\}
      }}
Note that  the summation range of  $k,j,p$ is now taken  over the original
      nondegenerate orbit \orbitrf. 
 We   can define a new integration variables $\tau \to \tau/2$
      for the first summand, $\tau \to 2\tau$ for the second and $\tau
      \to 2\tau+1$ for the third. Since the integration region is the
      entire upper half plane and the measure is invariant under this
      change  the integrals will remain unchanged  under
      this redefinition 
 \eqn\oneloopfourbc{\eqalign{I_{nd}&=\int_H{d^2\tau\over
      \tau_2} \sum_{A} {T_2\over \tau_2} \exp\Big(2\pi i 2k T -{\pi 2T_2\over \tau_2 U_2}\big| (1 U)
      \pmatrix{k&j\cr 0&p}
          \Big(\matrix{\tau \cr 1}\Big)\big|^2\Big)\cr
&\times \Big\{ QC_2(2\tau)+
      Q C_4({\tau\over 2})+Q C_4({\tau\over 2}-{1\over 2})\Big\} }}
we now have the remarkable identity \lerche\ 
\eqn\remid{QC_2(2\tau)+
      Q C_4({\tau\over 2})+Q C_4({\tau\over 2}-{1\over 2})= {1\over 2}}
 from which it follows
      that all the nontrivial dependence on powers of $e^{2\pi i \tau}$
      cancels out of \oneloopfourb\ and it  can be written as
\eqn\resind{\eqalign{I^{(2)+(3)+(4)}_{nd}&=2\int_{H}{d^2\tau\over
      \tau_2} \sum_{\quad k>0, 0\leq j<k, p\in Z}{T_2\over \tau_2}
       \exp\Big\{ 2\pi i 2 kpT
      -{\pi2 T_2\over \tau_2 U_2}\big| k\tau +j +pU\big|^2\Big\}\cr 
&= \sum_{N}{1\over 2}\sum_{N|m} e^{-2\pi N 2T}+cc. }}
Putting \unfoldedinc\  and \resind\  together  the result for
      $(\tr F^2)^2$ threshold is 
\eqn\finalreftw{I^{(\tr F^2)^2}_{nd}=\sum_N\Big({1\over 4}\sum_{N|m}{1\over m} e^{2\pi i
2NT}-{1\over 8}\sum_{N|m}{1\over m} e^{2\pi i 4NT}\Big)+cc.} 

Note that the identities \idenone\ and \remid\ imply that the
nonholomorphic part of the integrals which is produced by the presence
of the modular but not homolmorphic  $\hat{E}_2$ in \ctrftwo\ does not contribute to the threshold
integrals. This is rather remarkable and can be checked in detail, by expanding
\ctrftwo\ in powers of $1/\tau_2$. Using the same notation
$QC_k^{(i)},i=1,2$ for terms proportional to $1/(\tau_2)^i$   as in
\exptrftb, the vanishing of
these contributions is then a consequence of the following identities.
\eqn\indvana{\eqalign{QC_1^{(i)}-
    QC_2^{(i)}+QC_3^{(i)}+QC_4^{(i)}&=0,\quad  i=1,2\cr
\half QC_2^{(1)}(2\tau) + 2\Big( QC_4^{(1)}(\half
\tau)+QC_4^{(1)}(\half \tau-\half)\Big)&=0\cr
\quart QC_2^{(2)}(2\tau) + 4\Big( QC_4^{(2)}(\half
\tau)+QC_4^{(2)}(\half \tau-\half)\Big)&=0
}}
With the same rearrangement of the summation sectors as above.
\appendix{C}{Evaluation of the integrals}
In this appendix we review the evaluation of integrals appearing
in the heterotic threshold calculation. The basic technique was
developed in \dixon\ for more details in this context see
\costas\kiritsis.
\eqn\intega{I_m= T_2 \sum_{k>0, 0\leq j<k,  p\neq 0}e^{2\pi i  m kp T}\int {d^2\tau\over \tau_2^2}
\exp\Big(m {\pi T_2\over \tau_2 U_2}|k\tau+j+pU|^2\Big) } 
Where the parameter $m$ takes the values $m=2,4$ for the integrals
considered in Appendix B. Integrating over $\tau_1$ gives
\eqn\inetgb{I=\sum_{k>0, 0\leq j<k,  p\neq 0}{\sqrt{T_2U_2}\over
    k\sqrt{m}} e^{2\pi i m kp
T}\int {d\tau_2\over
\tau_2^{3/2}} e^{-{\pi T_2\over U_2}m k^2\tau_2
}e^{-\pi m p^2 T_2U_2/\tau_2}.}
 The integral over $\tau_2$ can be
done using the formula 
\eqn\itttwo{\int_0^\infty {dx\over
x^{3/2}} e^{-ax-b/x}=
\sqrt{{\pi\over b}}e^{-2\sqrt{ab}},}
 where 
\eqn\abdef{a= {\pi  T_2\over
U_2}m k^2,\quad b= \pi m p^2 T_2U_2.}

The final result  is the given by
\eqn\finintn{\eqalign{I_m&={1\over m} \sum_{0\leq j<k}\sum_{k>0,p>0}{1\over
    k|p|} e^{2\pi i m kp T}+cc.\cr&={1\over m}
  \sum_{N}\sum_{N|n}{1\over n}e^{2\pi i m NT}+cc.  }}

\listrefs
\end